\documentstyle[12pt]{article}
\input amssym.def
\input amssym.tex

\newcommand{\wiq}{\widetilde{q}}
\newcommand{\wir}{\widetilde{r}}
\newcommand{\whq}{\widehat{q}}

\newcommand{\bq}{{\bf q}}
\newcommand{\br}{{\bf r}}
\newcommand{\wbq}{\widetilde{\bf q}}
\newcommand{\wbr}{\widetilde{\bf r}}
\newcommand{\bA}{\mbox{\boldmath$A$}}
\newcommand{\bB}{\mbox{\boldmath$B$}}
\newcommand{\bC}{\mbox{\boldmath$C$}}
\newcommand{\bD}{\mbox{\boldmath$D$}}
\newcommand{\rA}{{\bf A}}
\newcommand{\rB}{{\bf B}}
\newcommand{\rC}{{\bf C}}
\newcommand{\rD}{{\bf D}}

\newcommand{\qed}{\rule{3mm}{3mm}}

\begin{document}

\renewcommand{\theequation}{\thesection.\arabic{equation}}

%
%

\begin{titlepage}
 {\LARGE
  \begin{center} 
On an integrable discretization \\ of the modified Korteweg-de Vries
 equation 
  \end{center}
 }

\vspace{1.5cm}

\begin{flushleft}{\large Yuri B. SURIS}\end{flushleft} \vspace{1.0cm}
Centre for Complex Systems and Visualization, University of Bremen,\\
Universit\"atsallee 29, 28359 Bremen, Germany\\
e-mail: suris @ cevis.uni-bremen.de 

\vspace{2.0cm}

{\small {\bf Abstract.} We find time discretizations for the two ''second
flows'' of the Ablowitz--Ladik hierachy. These discretizations are
described by local equations of motion, as opposed to the previously
known ones, due to Taha and Ablowitz. Certain superpositions of our maps
allow a one--field reduction and serve therefore as valid space--time
discretizations of the modified Korteweg-de Vries equation. We expect 
the performance of these discretizations to be much better then that of
the Taha--Ablowitz scheme. The way of finding interpolating Hamiltonians for 
our maps is also indicated, as well as the solution of an initial value 
problem in terms of matrix factorizations.}
\end{titlepage}

\setcounter{equation}{0}
\section{Introduction}
Already in the early period of the soliton theory it was realised that
by the numerical simulation of soliton equations it is highly desirable
that the difference schemes inherit the integrability property \cite{AL1},
\cite{AL2}, \cite{AT3}. These papers contained a full (space and time) 
discretization of such soliton equations as the nonlinear Schr\"odinger equation
(NLS), Korteweg--de Vries equation (KdV) and the modified Korteweg--de Vries
equation (mKdV). This was done in two steps. On the first step the space
discretization was performed \cite{AL1}. To this end the authors discretized
the auxiliary linear problem related to the corresponding soliton equation.
Concerning as an example the mKdV equation,
\begin{equation}\label{MKDV}
q_t=q_{xxx}\mp 6q_xq^2
\end{equation}
the associated linear problem is the Zakharov--Shabat problem
\begin{equation}\label{lin pr}
\Psi_x=\left(\begin{array}{cc}i\zeta & q \\ \\
\pm q & -i\zeta\end{array}\right)\Psi.
\end{equation}
Its discretization chosen in \cite{AL1} is
\begin{equation}\label{d lin pr}
\Psi_{k+1}=\left(\begin{array}{cc} \lambda & q_k \\ \\
\pm q_k & \lambda^{-1}\end{array}\right)\Psi_k,
\end{equation}
and the space discretization of (\ref{MKDV}) found in \cite{AL1} is
\begin{eqnarray}\label{dMKDV}
\dot{q}_k&=&(1\mp q_k^2)\Big(q_{k+2}-2q_{k+1}+2q_{k-1}-q_{k-2}\nonumber\\
&&\mp\,q_{k+1}(q_{k+2}q_{k+1}+q_{k+1}q_k)
\pm q_{k-1}(q_{k}q_{k-1}+q_{k-1}q_{k-2})\Big)
\end{eqnarray}
(to perform the corresponding continuous limit, one has to assume
in (\ref{dMKDV}) $q_k=\epsilon q(k\epsilon)$, to rescale the time
$t\mapsto t/(2\epsilon^3)$, and then to send $\epsilon\to 0$).

The second step of discretization -- the time discretization -- was performed 
in \cite{AL2} (for NLS) and in \cite{AT3} (for KdV and mKdV). The approach to 
this second step of the discretization process was fundamentally different: 
the linear problem (\ref{d lin pr}) was not modified any more, and only a
suitable choice of a (discrete)--time evolution of the wave function
$\Psi_k$ was imposed. The outcome of \cite{AT3} was an excellent numerical
scheme for the mKdV, though not free of some drawbacks, the main one being
the non-locality (which means that $\wiq_k$, the discrete time update of
$q_k$, depends explicitly on all $q_j$'s and $\wiq_j$'s with $j\leq k$). 
This property 
is unpleasant from the aesthetical point of view as well as from the practical
one (since it implies a large amount of computations by numerical realization).

In a more modern language, time discretizations in \cite{AL2}, \cite{AT3}
were sought in the same hierarchies to which the underlying continuous time 
systems belong. Recently this approach
was pushed forward as a systematic procedure of obtaining integrable 
discretizations \cite{S1}--\cite{S4}. The results of \cite{AL2} were 
re-considered in \cite{S5}, where a significant amendment was achieved. 
Namely, the non-locality of the schemes in \cite{AL2} for NLS was overcome. 
In the present paper we do an analogous work 
for the time discretization of (\ref{dMKDV}) from \cite{AT3}.

\setcounter{equation}{0}
\section{Ablowitz--Ladik hierarchy}

To deal with the equation (\ref{MKDV}) in a slightly more systematic way, 
one has to consider it as a particular case of the following, more general 
system:
\begin{equation}\label{cMKDV}
q_t=q_{xxx}-6q_xqr,\qquad r_t=r_{xxx}-6qr_xr
\end{equation}
under the reduction 
\begin{equation}\label{red}
r=\pm q.
\end{equation}

Analogously, the space discretization (\ref{dMKDV}) arises by the same
reduction from the more general system
\begin{eqnarray}\label{AL2}
\dot{q}_k & = & (1-q_kr_k)\Big(q_{k+2}-2q_{k+1}+2q_{k-1}-q_{k-2}\nonumber\\
&&-q_{k+1}(q_{k+2}r_{k+1}+q_{k+1}r_k+q_kr_{k-1})\nonumber\\
&&+q_{k-1}(q_{k-2}r_{k-1}+q_{k-1}r_k+q_kr_{k+1})\Big)\nonumber\\ \\
\dot{r}_k & = & (1-q_kr_k)\Big(r_{k+2}-2r_{k+1}+2r_{k-1}-r_{k-2}\nonumber\\
&&-r_{k+1}(q_{k+1}r_{k+2}+q_kr_{k+1}+q_{k-1}r_k)\nonumber\\
&&+r_{k-1}(q_{k-1}r_{k-2}+q_kr_{k-1}+q_{k+1}r_k)\Big)\nonumber
\end{eqnarray}
Below we consider this system either on an infinite lattice 
($k\in {\Bbb Z}$) under the boundary conditions of a rapid decay 
($|q_k|, |r_k|\to 0$ as $k\to\pm\infty$),
or on a finite lattice ($1\le k\le N$) under the periodic boundary conditions 
($q_0\equiv q_N$, $r_0\equiv r_N$, $q_{N+1}\equiv q_1$, $r_{N+1}\equiv r_1$).
In any case we denote by $\bq$ ($\br$) the (infinite- or 
finite-dimensional) vector with the components $q_k$ (resp. $r_k$).

From the modern point of view, the system (\ref{AL2}) is
a representative of a whole hierarchy of commuting Hamiltonian flows, the
Ablowitz--Ladik hierarchy.
An object of the principle importance in the description of this hierarchy
is the $2\times 2$ Lax matrix
\begin{equation}\label{AL Lk}
L_k=L_k(\bq,\br)=\left(
  \begin{array}{cc} \lambda & q_k \\ \\ r_k & \lambda^{-1}\end{array}
     \right)
\end{equation}
depending on the variables $\bq, \br$ and on the additional (spectral) 
parameter $\lambda$. Each flow of the hierarchy allows a commutation
representation (semi-discrete version of a zero--curvature representation)
\begin{equation}\label{AL Lax}
\dot{L}_k=M_{k+1}L_k-L_kM_k
\end{equation}
with some $2\times 2$ matrix $M_k=M_k(\bq,\br,\lambda)$. 

The Hamiltonians of the commuting flows are the coefficients in the Laurent
expansion of the trace ${\rm tr}\; T_N(\bq,\br,\lambda)$ where $T_N$ is the 
monodromy matrix
\begin{equation}\label{AL monodromy}
T_N=L_N\cdot L_{N-1}\cdot\ldots\cdot L_2\cdot L_1,
\end{equation}
supplied by the function
\begin{equation}\label{AL H0}
H_0(\bq,\br)=\log\det T_N=\sum_{k=1}^N\log(1-q_kr_k).
\end{equation}
The Poisson bracket on the phase space is given by
\begin{equation}\label{AL PB}
\{q_k,r_j\}=(1-q_kr_k)\delta_{jk},\quad \{q_k,q_j\}=\{r_k,r_j\}=0.
\end{equation}

The involutivity of all integrals of motion follows from the fundamental
$r$--matrix relation:
\begin{equation}\label{r mat PB}
\{L(\lambda)\stackrel{\bigotimes}{,}L(\mu)\}=
[L(\lambda)\otimes L(\mu),\rho(\lambda,\mu)],
\end{equation}
where
\begin{equation}\label{AL r mat}
\rho(\lambda,\mu)=\left(\begin{array}{cccc}
\frac{1}{2}\frac{\lambda^2+\mu^2}{\lambda^2-\mu^2} & 0 & 0 & 0 \\
0 & \frac{1}{2} & \frac{\lambda\mu}{\lambda^2-\mu^2} & 0 \\
0 & \frac{\lambda\mu}{\lambda^2-\mu^2} & -\frac{1}{2} & 0 \\
0 & 0 & 0 & \frac{1}{2}\frac{\lambda^2+\mu^2}{\lambda^2-\mu^2}
\end{array}\right).
\end{equation}

We refer the reader to \cite{FT} for the general formula expressing the matrix
$M_k$ in terms of the corresponding Hamiltonian function $H(\bq,\br)$ and the
$r$--matrix $\rho(\lambda,\mu)$.

It is easy to see that the following functions belong to the involutive
family generated by ${\rm tr}\; T_N$:
\begin{equation}\label{AL H pm1}
H_1(\bq,\br)=\sum_{k=1}^N q_{k+1}r_k, \quad 
H_{-1}(\bq,\br)=\sum_{k=1}^N q_kr_{k+1},
\end{equation}
\begin{eqnarray}
H_2(\bq,\br)&=&\sum_{k=1}^N q_{k+2}r_k-\sum_{k=1}^N q_{k+2}q_{k+1}r_{k+1}r_k-
\frac{1}{2}\sum_{k=1}^Nq_{k+1}^2r_k^2,\label{AL H2}\\
H_{-2}(\bq,\br)&=&\sum_{k=1}^N q_kr_{k+2}-\sum_{k=1}^N q_kq_{k+1}r_{k+1}r_{k+2}-
\frac{1}{2}\sum_{k=1}^Nq_k^2r_{k+1}^2.\label{AL h-2}
\end{eqnarray}

The corresponding Hamiltonian flows are described by the differential equations
\begin{equation}\label{AL flow+1}
{\cal F}_1:\;\left\{\begin{array}{l}
\dot{q}_k=q_{k+1}(1-q_kr_k)\\ 
\dot{r}_k=-r_{k-1}(1-q_kr_k)
\end{array}\right.
\end{equation}
\begin{equation}\label{AL flow-1}
{\cal F}_{-1}:\;\left\{\begin{array}{l}
\dot{q}_k=q_{k-1}(1-q_kr_k)\\
\dot{r}_k=-r_{k+1}(1-q_kr_k)
\end{array}\right.
\end{equation}
\begin{equation}\label{AL flow+2}
{\cal F}_2:\; \left\{\begin{array}{l}
\dot{q}_k=\Big(q_{k+2}-q_{k+1}(q_{k+2}r_{k+1}
+q_{k+1}r_k+q_kr_{k-1})\Big)\,(1-q_kr_k)\\ 
\dot{r}_k=-\Big(r_{k-2}-r_{k-1}(q_{k-1}r_{k-2}
+q_kr_{k-1}+q_{k+1}r_k)\Big)\,(1-q_kr_k)
\end{array}\right.
\end{equation}
\begin{equation}\label{AL flow-2}
{\cal F}_{-2}:\; \left\{\begin{array}{l}
\dot{q}_k=\Big(q_{k-2}-q_{k-1}(q_{k-2}r_{k-1}
+q_{k-1}r_k+q_kr_{k+1})\Big)\,(1-q_kr_k)\\ 
\dot{r}_k=-\Big(r_{k+2}-r_{k+1}(q_{k+1}r_{k+2}
+q_kr_{k+1}+q_{k-1}r_k)\Big)\,(1-q_kr_k)
\end{array}\right.
\end{equation}

The  flow (\ref{AL2}) is an obvious superposition of these
more fundamental and simple flows, namely
\[
{\cal F}_2(t)\circ{\cal F}_1(-2t)\circ {\cal F}_{-2}(-t)\circ {\cal F}_{-1}(2t)
\]
In what follows we will concentrate on the flows
\begin{equation}\label{AL+2 flow}
{\cal F}_2(t)\circ{\cal F}_1(-ct)
\end{equation}
\begin{equation}\label{AL-2 flow}
{\cal F}_{-2}(t)\circ {\cal F}_{-1}(-ct),
\end{equation}
having in mind that in the application to the mKdV case the value $c=2$ is of
interest.

The matrix $M_k$ corresponding to the flow (\ref{AL+2 flow})
is given by the formula
\begin{equation}\label{AL M+2}
M^{(2)}_k(c)=\left(\begin{array}{cc}
\lambda^4-\lambda^2{\cal A}_k^{(2)}-{\cal A}_k^{(0)} & 
\lambda^3 q_k+\lambda{\cal B}_k^{(1)}\\ \\ 
\lambda^3 r_{k-1}+\lambda{\cal C}_k^{(1)} & 
\lambda^2 q_kr_{k-1} \end{array}\right)
\end{equation}
with
\begin{eqnarray*}
{\cal A}_k^{(2)}&=&c+q_kr_{k-1}\\
{\cal B}_k^{(1)}&=&
q_{k+1}-q_k(c+q_{k+1}r_k+q_kr_{k-1})\\
{\cal C}_k^{(1)}&=&
r_{k-2}-r_{k-1}(c+q_{k-1}r_{k-2}+q_kr_{k-1})\\
{\cal A}_k^{(0)}&=&
q_{k+1}r_{k-1}+q_kr_{k-2}
-q_kr_{k-1}(c+q_{k+1}r_k+q_kr_{k-1}+q_{k-1}r_{k-2}).
\end{eqnarray*}

Similarly, the matrix $M_k$ corresponding to the flow (\ref{AL-2 flow})
is given by
\begin{equation}\label{AL M-2}
M^{(-2)}_k(c)=\left(\begin{array}{cc}
-\lambda^{-2}q_{k-1}r_k &
-\lambda^{-3}q_{k-1}-\lambda^{-1}{\cal B}_k^{(-1)}\\ \\ 
-\lambda^{-3}r_k-\lambda^{-1}{\cal C}_k^{(-1)}&
 -\lambda^{-4}+\lambda^{-2}{\cal D}_k^{(-2)}+{\cal D}_k^{(0)}
\end{array}\right)
\end{equation}
with
\begin{eqnarray*}
{\cal D}_k^{(-2)} & = & c+q_{k-1}r_k\\
{\cal B}_k^{(-1)}&=&
q_{k-2}-q_{k-1}(c+q_{k-2}r_{k-1}+q_{k-1}r_k)\\
{\cal C}_k^{(-1)}&=&
r_{k+1}-r_k(c+q_kr_{k+1}+q_{k-1}r_k)\\
{\cal D}_k^{(0)}&=&
q_{k-2}r_k+q_{k-1}r_{k+1}
-q_{k-1}r_k(c+q_{k-2}r_{k-1}+q_{k-1}r_k+q_kr_{k+1}).
\end{eqnarray*}

Obviously, one has:
\[
M_k^{(2)}(c)=M_k^{(2)}(0)-c M_k^{(1)},\quad
M_k^{(-2)}(c)=M_k^{(-2)}(0)-c M_k^{(-1)},
\]
where the matrices 
\begin{equation}\label{AL M+1}
M_k^{(1)}=\left(\begin{array}{cc}
\lambda^2-q_kr_{k-1} & \lambda q_k\\ \\ 
\lambda r_{k-1} & 0 \end{array}\right),
\end{equation}
\begin{equation}\label{AL M-1}
M_k^{(-1)}=\left(\begin{array}{cc}
0 & -\lambda^{-1}q_{k-1}\\ \\ 
-\lambda^{-1}r_k & -\lambda^{-2}+q_{k-1}r_k
\end{array}\right)
\end{equation}
correspond to the flows ${\cal F}_{\pm 1}$.

The matrix $M_k$ for the system (\ref{AL2}) is equal to
\[
M_k=M_k^{(2)}(2)-M_k^{(-2)}(2).
\]

\setcounter{equation}{0}
\section{General remarks about the time discretization}

In \cite{AT3} Taha and Ablowitz constructed a time discretizations of the 
system (\ref{AL2}), thus achieving a full discretization of the system 
(\ref{cMKDV}). The basic feature of the time discretization in \cite{AT3} 
is following: it admits a discrete analog of the zero--curvature 
representation,
\begin{equation}\label{d zero curv}
\widetilde{L}_kV_k=V_{k+1}L_k
\end{equation}
with the same matrix $L_k$ as the underlying continuous time system. (In 
(\ref{d zero curv}) and below we use the tilde to denote the $h$--shift
in the discrete time $h{\Bbb Z}$). In a more modern language, the maps
generated by the discretizations in \cite{AT3} belong to the same integrable 
hierarchy as the continuous time system (\ref{AL2}).

We shall not need concrete expressions for the entries of the matrix
$V_k$ and the corresponding evolution equations obtained in \cite{AT3}.
However, the details of the derivation seem to be never published, therefore 
we present them (in a slightly amended form) in the Appendix. Here we restrict
ourselves with some general remarks following from the work of Taha and Ablowitz
and necessary for the following presentation.

They considered difference equations allowing a commutation representation
(\ref{d zero curv}) with the matrix $L_k$ given by (\ref{AL Lk}) and assumed 
that the entries of the matrix 
\begin{equation}\label{dAL Vk}
V_k=\left(\begin{array}{cc}A_k & B_k\\ \\C_k & D_k\end{array}\right)
\end{equation}
have the following $\lambda$--dependence: 
\begin{eqnarray*}
A_k(\lambda) & = & 1+h\lambda^4A_k^{(4)}+h\lambda^2A_k^{(2)}+hA_k^{(0)}+
h\lambda^{-2}A_k^{(-2)}+h\lambda^{-4}A_k^{(-4)}\\
D_k(\lambda) & = & 1+h\lambda^4D_k^{(4)}+h\lambda^2D_k^{(2)}+hD_k^{(0)}+
h\lambda^{-2}D_k^{(-2)}+h\lambda^{-4}D_k^{(-4)}\\
B_k(\lambda) & = & h\lambda^{3}B_k^{(3)}+h\lambda B_k^{(1)}
+h\lambda^{-1}B_k^{(-1)}+h\lambda^{-3}B_k^{(-3)}\\
C_k(\lambda) & = & h\lambda^{3}C_k^{(3)}+h\lambda C_k^{(1)}
+h\lambda^{-1}C_k^{(-1)}+h\lambda^{-3}C_k^{(-3)}.
\end{eqnarray*}

They showed that each such difference equation may be completely
characterized (in the case of rapidly decaying boundary conditions) by the 
limit values
\begin{equation}\label{dAL2 aldel}
\alpha^{(j)}=\lim_{k\to\pm\infty} A_k^{(j)},\quad
\delta^{(j)}=\lim_{k\to\pm\infty} D_k^{(j)} \quad (j=4,2,0,-2,-4)
\end{equation}

Under the condition
\begin{equation}\label{dAL2 al=del}
\alpha^{(j)}=\delta^{(-j)} \quad (j=4,2,0,-2,-4)
\end{equation}
the corresponding difference equation allows the reduction
\begin{equation}\label{dAL2 red}
\br=\pm\bq.
\end{equation}

The last statement is easy to see directly. Indeed, the discrete 
zero--curvature equation (\ref{d zero curv}) is equivalent to
the following four equations:
\begin{eqnarray}
A_{k+1}(\lambda)-A_k(\lambda) & = &
\lambda^{-1}\Big(\wiq_k C_k(\lambda)-r_k B_{k+1}(\lambda)\Big)
\label{dAL2 A}\\
D_{k+1}(\lambda)-D_k(\lambda) & = &
\lambda\Big(\wir_k B_k(\lambda)-q_kC_{k+1}(\lambda)\Big)
\label{dAL2 D}\\
\lambda^{-1}B_{k+1}(\lambda)-\lambda B_k(\lambda) & = &
\wiq_k D_k(\lambda)-q_k A_{k+1}(\lambda)
\label{dAL2 B}\\
\lambda C_{k+1}(\lambda)-\lambda^{-1}C_k(\lambda) & = &
\wir_k A_k(\lambda)-r_k D_{k+1}(\lambda)
\label{dAL2 C}
\end{eqnarray}
Obviously, these equations allow the reduction
\[
\br=\pm\bq,\quad A(\lambda)=D(\lambda^{-1}),\quad 
B(\lambda)=\pm C(\lambda^{-1}).
\]
But in this reduction, obviously, the condition (\ref{dAL2 al=del}) is
satisfied. Since the difference equation is completely characterised by the
quantities (\ref{dAL2 aldel}), the condition (\ref{dAL2 al=del}) is also
sufficient for the above reduction to be admissible.

Equating coefficients by different powers of $\lambda$ in 
(\ref{dAL2 A})--(\ref{dAL2 C}), one obtains 20 equations. In \cite{AT3}, 
Taha and Ablowitz assumed a special role of evolution equations
for $q_k$, $r_k$ to two of these equations, using the other 18 to determine 
the 18 unknown functions
$A_k^{(j)}$, $D_k^{(j)}$ ($j=4,2,0,-2,-4$), and $B_k^{(j)}$, $C_k^{(j)}$
($j=3,1,-1,-3$). This way of dealing with the problem (inherited from the 
continuous time case) resulted in highly
nonlocal expressions. This feature of the resulting difference scheme made
its numerical realization extremely time consuming. Nevertheless, even 
despite the drawback of non-locality, this difference scheme
proved to be the best among the numerical methods tested in \cite{AT3}.
The reason for this lies undoubtedly in the integrable nature of this
scheme.

The goal of the present paper is to demonstrate how the feature of nonlocality
may be overcome. This is achieved with the help of two basic ideas. First,
we factorize the difference scheme into the product of several simpler ones,
corresponding to the fundamental and simple flows of the Ablowitz--Ladik
hierarchy. Second, by dealing with these simpler schemes we consider all
the 20 equations on an equal footing. This allows us to derive simple local
schemes, which we expect to exceed the original one due to Taha
and Ablowitz.

\setcounter{equation}{0}
\section{Local discretizations of the flows ${\cal F}_{\pm 2}$}
In the present Section we introduce two maps which serve as time discretizations
of the flows (\ref{AL+2 flow}), (\ref{AL-2 flow}).

{\bf Theorem 1.} {\it Consider the map ${\cal T}_{2}(h;c): (\bq,\br)
\mapsto (\wbq,\wbr)$ defined by the following equations of motion:
\begin{equation}\label{dAL2}
{\cal T}_{2}(h;c):\left\{\begin{array}{l}
(\wiq_k-q_k)/h=
\displaystyle\frac
{(q_{k+2}-q_{k+1}P_{k+1})\,(1-q_k\wir_k)}{(1-hq_{k+1}\wir_{k-1})},\\ \\
(\wir_k-r_k)/h=
-\displaystyle\frac
{(\wir_{k-2}-\wir_{k-1}P_{k})\,(1-q_k\wir_k)}{(1-hq_{k+1}\wir_{k-1})}
\end{array}\right.
\end{equation}
where  the {\rm local} quantity $P_k=P(q_{k+1},q_{k},q_{k-1},\wir_{k},
\wir_{k-1},\wir_{k-2};h;c)$ is defined by
\[
\Big(P_k-hq_{k+1}\wir_{k-2}\Big)\,\frac{(1-hq_kq_{k-1}\wir_{k-1}\wir_{k-2})\,
(1-hq_{k+1}q_k\wir_k\wir_{k-1})}{(1-hq_k\wir_{k-2})\,(1-hq_{k+1}\wir_{k-1})}
\]
\begin{equation}\label{dAL2 Pk}
-\Big(q_{k+1}\wir_k+q_k\wir_{k-1}+q_{k-1}\wir_{k-2}
-hq_{k+1}q_kq_{k-1}\wir_k\wir_{k-1}\wir_{k-2}\Big)=c,
\end{equation}
so that
\[
P_k=c+q_{k+1}\wir_k+q_k\wir_{k-1}+q_{k-1}\wir_{k-2}+O(h).
\]
These equations approximate the flow {\rm(\ref{AL+2 flow})} and have the 
commutation representation {\rm (\ref{d zero curv})} with the matrix
\begin{equation}\label{dAL2 Vk}
V_k^{(2)}=
\left(\begin{array}{cc}
1+h\lambda^4-h\lambda^{2}A_k^{(2)}-hA_k^{(0)} & 
h\lambda^{3}q_{k}+h\lambda B_k^{(1)}\\ \\
h\lambda^{3}\wir_{k-1}+h\lambda C_k^{(1)} & 
1+h\lambda^{2}q_k\wir_{k-1}\end{array}\right).
\end{equation}
Here}
\begin{eqnarray}
A_k^{(2)} & = & P_k-q_{k+1}r_k-\wiq_{k-1}\wir_{k-2}
\label{dAL2 A2}\\
B_k^{(1)} & = & q_{k+1}-q_k(P_k-\wiq_{k-1}\wir_{k-2})
\label{dAL2 B1}\\
C_k^{(1)} & = & \wir_{k-2}-\wir_{k-1}(P_k-q_{k+1}r_k)
\label{dAL2 C1}\\
A_k^{(0)}  & = & q_{k+1}\wir_{k-1}+q_k\wir_{k-2}-q_k\wir_{k-1}P_k
\label{dAL2 A0}
\end{eqnarray}

{\bf Theorem 2.} {\it Consider the map ${\cal T}_{-2}(h;c): (\bq,\br)
\mapsto (\wbq,\wbr)$ defined by the following equations of motion:
\begin{equation}\label{dAL-2}
{\cal T}_{-2}(h;c):\left\{\begin{array}{l}
(\wiq_k-q_k)/h=
\displaystyle\frac
{(\wiq_{k-2}-\wiq_{k-1}S_k)\,(1-\wiq_kr_k)}{(1+h\wiq_{k-1}r_{k+1})},\\ \\
(\wir_k-r_k)/h=
-\displaystyle\frac
{(r_{k+2}-r_{k+1}S_{k+1})\,(1-\wiq_kr_k)}{(1+h\wiq_{k-1}r_{k+1})}
\end{array}\right.
\end{equation}
where  the {\rm local} quantity $S_k=S(\wiq_{k-2},\wiq_{k-1},\wiq_k,r_{k-1},
r_k,r_{k+1};h;c)$ is defined by
\[
\Big(S_k+h\wiq_{k-2}r_{k+1}\Big)\,\frac{(1+h\wiq_{k-2}\wiq_{k-1}r_{k-1}r_k)\,
(1+h\wiq_{k-1}\wiq_kr_kr_{k+1})}{(1+h\wiq_{k-2}r_k)\,(1+h\wiq_{k-1}r_{k+1})}
\]
\begin{equation}\label{dAL-2 Sk}
-\Big(\wiq_kr_{k+1}+\wiq_{k-1}r_k+\wiq_{k-2}r_{k-1}+h\wiq_{k-2}\wiq_{k-1}\wiq_k
r_{k-1}r_kr_{k+1}\Big)=c,
\end{equation}
so that
\[
S_k=c+\wiq_kr_{k+1}+\wiq_{k-1}r_k+\wiq_{k-2}r_{k-1}+O(h).
\]
These equations approximate the flow {\rm(\ref{AL-2 flow})}
and have the commutation representation {\rm (\ref{d zero curv})} with 
the matrix
\begin{equation}\label{dAL-2 Vk}
V_k^{(-2)}=
\left(\begin{array}{cc}
1-h\lambda^{-2}\wiq_{k-1}r_k & -h\lambda^{-3}\wiq_{k-1}-h\lambda^{-1}B_k^{(-1)}\\ \\
-h\lambda^{-3}r_k-h\lambda^{-1} C_k^{(-1)} & 
1-h\lambda^{-4}+h\lambda^{-2}D_k^{(-2)}+hD_k^{(0)}\end{array}\right).
\end{equation}
Here}
\begin{eqnarray}
D_k^{(-2)} & = & S_k-q_kr_{k+1}-\wiq_{k-2}\wir_{k-1}
\label{dAL-2 D-2}\\
B_k^{(-1)} & = & \wiq_{k-2}-\wiq_{k-1}(S_k-q_kr_{k+1})
\label{dAL-2 B-1}\\
C_k^{(-1)} & = & r_{k+1}-r_k(S_k-\wiq_{k-2}\wir_{k-1})
\label{dAL-2 C-1}\\
D_k^{(0)}  & = & \wiq_{k-1}r_{k+1}+\wiq_{k-2}r_k-\wiq_{k-1}r_kS_k
\label{dAL-2 D0}
\end{eqnarray}

{\bf Proof.} Since the both Theorems are proved similarly, we give here
only the proof of the Theorem 2. Substituting the ansatz (\ref{dAL-2 Vk})
for the matrix $V_k^{(-2)}$ into (\ref{d zero curv}), one sees
that the following 6 equations have to be satisfied:
\begin{eqnarray}
D_{k+1}^{(-2)}-D_k^{(-2)} & = &q_kr_{k+1}-\wiq_{k-1}\wir_k
\label{dAL-2 proof D-2}\\
D_{k+1}^{(0)}-D_k^{(0)} & = & q_kC_{k+1}^{(-1)}-\wir_kB_k^{(-1)}
\label{dAL-2 proof D0}\\
B_{k+1}^{(-1)} & = & \wiq_{k-1}-\wiq_k
\left(q_{k}r_{k+1}+D_{k}^{(-2)}\right)
\label{dAL-2 proof B}\\
C_{k}^{(-1)} & = & r_{k+1}-r_k
\left(\wiq_{k-1}\wir_k+D_{k+1}^{(-2)}\right)
\label{dAL-2 proof C}\\
\wiq_k-q_k & = & hB_{k}^{(-1)}-h\wiq_kD_{k}^{(0)}
\label{dAL-2 proof q}\\
\wir_k-r_k & = &-hC_{k+1}^{(-1)}+hr_kD_{k+1}^{(0)}
\label{dAL-2 proof r}
\end{eqnarray}
As indicated above, the approach by Taha and Ablowitz to these equations would
be as follows: consider the last 2 equations as the equations of motion, 
where the (non-local) expressions for the quantities 
$B_k^{(-1)}$, $C_k^{(-1)}$, $D_k^{(0)}$ follow directly from the first 4 
equations. The crucial feature of our approach to the solution of these 
equations is that we use {\it all} 6 equations to derive {\it local} 
expressions. In other words, we do not assume that the last two of them play 
a special role.

Note first of all that due to (\ref{dAL-2 proof D-2}), the equations
(\ref{dAL-2 proof B}), (\ref{dAL-2 proof C}) may be equivalently re-written
as
\begin{eqnarray}
B_{k+1}^{(-1)} & = & \wiq_{k-1}-\wiq_k
\left(\wiq_{k-1}\wir_k+D_{k+1}^{(-2)}\right)
\label{dAL-2 proof B-1}\\
C_{k}^{(-1)} & = & r_{k+1}-r_k
\left(q_{k}r_{k+1}+D_{k}^{(-2)}\right)
\label{dAL-2 proof C-1}
\end{eqnarray}

Now we introduce the auxiliary quantity $S_k$ by the formula
\begin{equation}\label{dAL-2 proof Sk}
S_k=D_k^{(-2)}+q_kr_{k+1}+\wiq_{k-2}\wir_{k-1}.
\end{equation}
which immediately gives (\ref{dAL-2 D-2}). With the help of this 
quantity we re-write the equations (\ref{dAL-2 proof B-1}), 
(\ref{dAL-2 proof C-1}) as (\ref{dAL-2 B-1}) and (\ref{dAL-2 C-1}), 
respectively. Substituting (\ref{dAL-2 B-1}), (\ref{dAL-2 C-1}) into
(\ref{dAL-2 proof D0}), we obtain the equation which may be written as
\begin{eqnarray}
D_{k+1}^{(0)}-D_k^{(0)} & = &
\wiq_kr_{k+2}-\wiq_{k-2}r_k-\wiq_kr_{k+1}S_{k+1}=\wiq_{k-1}r_kS_k
\nonumber\\
&&-(\wiq_k-q_k)(r_{k+2}-r_{k+1}S_{k+1})-(\wir_k-r_k)(\wiq_{k-2}-\wiq_{k-1}S_k).
\label{dAL-2 proof aux1}
\end{eqnarray}
Similarly, substituting (\ref{dAL-2 B-1}), (\ref{dAL-2 C-1}) into
(\ref{dAL-2 proof q}), (\ref{dAL-2 proof r}) we obtain two equations which may
be put down as 
\begin{equation}\label{dAL-2 proof aux2}
(1+h\wiq_{k-1}r_{k+1})\,(\wiq_k-q_k)=h(\wiq_{k-2}-\wiq_{k-1}S_k)
-h\wiq_k\left(D_k^{(0)}-\wiq_{k-1}r_{k+1}\right)
\end{equation}
\begin{equation}\label{dAL-2 proof aux3}
(1+h\wiq_{k-1}r_{k+1})\,(\wir_k-r_k)=-h(r_{k+2}-r_{k+1}S_{k+1})
+hr_k\left(D_{k+1}^{(0)}-\wiq_{k-1}r_{k+1}\right)
\end{equation}
Multiplying (\ref{dAL-2 proof aux1}) by $(1+h\wiq_{k-1}r_{k+1})$ and using
on the right--hand side the expressions (\ref{dAL-2 proof aux2}),
(\ref{dAL-2 proof aux3}), we obtain an equation, which after some identical
re-arrangements may be presented as
\[
\left(1+hD_{k+1}^{(0)}\right)
\Big(1+h\wiq_{k-1}r_{k+1}+h\wiq_{k-2}r_{k}-h\wiq_{k-1}r_{k}S_{k}\Big)
\]
\[
=\left(1+hD_{k}^{(0)}\right)
\Big(1+h\wiq_kr_{k+2}+h\wiq_{k-1}r_{k+1}-h\wiq_{k}r_{k+1}S_{k+1}\Big).
\]
So, the quantity
\[
\left(1+hD_{k}^{(0)}\right)/
\Big(1+h\wiq_{k-1}r_{k+1}+h\wiq_{k-2}r_{k}-h\wiq_{k-1}r_{k}S_{k}\Big)
\]
does not depend on $k$, and, taking the limit $k\to\pm\infty$,
we see that it is equal to 1, which gives (\ref{dAL-2 D0}).
Substituting this result back in (\ref{dAL-2 proof aux2}),
(\ref{dAL-2 proof aux3}), we obtain (\ref{dAL-2}).

Now to finish the proof it remains to obtain the local expression 
(\ref{dAL-2 Sk}). To this end we first of all re-write the equation  
(\ref{dAL-2 proof D-2}) in terms of $S_k$ and find the following difference
equation:
\begin{equation}\label{dAL-2 S recur}
S_{k+1}-S_k=q_{k+1}r_{k+2}-\wiq_{k-2}\wir_{k-1}.
\end{equation}
As it stands, it does not allow a local solution. In order to achieve our goal,
we have to once more switch on the equations of motion (\ref{dAL-2}).
Putting on the right--hand side of (\ref{dAL-2 S recur})
\[
q_{k+1}=\wiq_{k+1}-h\,\frac{(\wiq_{k-1}-\wiq_kS_{k+1})(1-\wiq_{k+1}r_{k+1})}
{(1+h\wiq_kr_{k+2})},
\]
\[
\wir_{k-1}=r_{k-1}-h\,\frac{(r_{k+1}-r_kS_k)(1-\wiq_{k-1}r_{k-1})}
{(1+h\wiq_{k-2}r_k)},
\]
we obtain after some re-arrangements:
\[
\Big(S_{k+1}+h\wiq_{k-1}r_{k+2}\Big)\,
\frac{(1+h\wiq_k\wiq_{k+1}r_{k+1}r_{k+2})}{(1+h\wiq_kr_{k+2})}
-\Big(S_{k}+h\wiq_{k-2}r_{k+1}\Big)\,
\frac{(1+h\wiq_{k-2}\wiq_{k-1}r_{k-1}r_{k})}{(1+h\wiq_{k-2}r_{k})}
\]
\[
=(\wiq_{k+1}r_{k+2}-\wiq_{k-2}r_{k-1})(1+h\wiq_{k-1}r_{k+1}).
\]
Multiplying both sides by
\[
\frac{(1+h\wiq_{k-1}\wiq_kr_kr_{k+1})}{(1+h\wiq_{k-1}r_{k+1})},
\]
we arrive at the identity which may be read in the following way: the 
left--hand side of (\ref{dAL-2 Sk}) does not depend on $k$. The value of
this constant may be determined from the $k\to\pm\infty$ limit. \qed

{\bf Remark 1.} It might be desirable to have the expressions for all entries 
of the matrix $V_k^{(2)}$ in terms of $(\bq,\wbr)$ solely, and of
the matrix $V_k^{(-2)}$  -- in terms of $(\wbq,\br)$. Notice that
the quantities $A_k^{(0)}$, $D_k^{(0)}$ already have the desired form.
For the quantities $B_k^{(1)}$, $C_k^{(1)}$ it is possible to derive from 
(\ref{dAL2 Pk}), (\ref{dAL2}) the following nice expressions:
\begin{equation}
P_k-q_{k+1}r_k=(c+q_k\wir_{k-1}+q_{k-1}\wir_{k-2})\,
\frac{(1-hq_k\wir_{k-2})}{(1-hq_kq_{k-1}\wir_{k-1}\wir_{k-2})},
\end{equation}
\begin{equation}
P_k-\wiq_{k-1}\wir_{k-2}=(c+q_{k+1}\wir_k+q_k\wir_{k-1})\,
\frac{(1-hq_{k+1}\wir_{k-1})}{(1-hq_{k+1}q_k\wir_k\wir_{k-1})}.
\end{equation}
Analogously, for the quantities $B_k^{(-1)}$, $C_k^{(-1)}$ it follows from
(\ref{dAL-2 Sk}), (\ref{dAL-2}):
\begin{equation}
S_k-q_kr_{k+1}=(c+\wiq_{k-1}r_k+\wiq_{k-2}r_{k-1})\,
\frac{(1+h\wiq_{k-2}r_k)}{(1+h\wiq_{k-2}\wiq_{k-1}r_{k-1}r_k)},
\end{equation}
\begin{equation}
S_k-\wiq_{k-2}\wir_{k-1}=(c+\wiq_kr_{k+1}+\wiq_{k-1}r_k)\,
\frac{(1+h\wiq_{k-1}r_{k+1})}{(1+h\wiq_{k-1}\wiq_kr_kr_{k+1})}.
\end{equation}
Unfortunately, analogous expressions for $A_k^{(2)}$, resp. $D_k^{(-2)}$ 
are complicated and non-elegant.

{\bf Remark 2.} Notice that from the Theorems 1,2 we can recover the maps
from the previous paper \cite{S5}, together with the corresponding matrices
$V_k$. Indeed, rescaling $h\mapsto h/c$ and then sending $c\to\infty$, we
find from the Theorem 1 the map
\begin{equation}\label{dAL1}
{\cal T}_1(-h):\left\{\begin{array}{l}
(\wiq_k-q_k)/h=-q_{k+1}(1-q_k\wir_k),\\ \\
(\wir_k-r_k)/h=\wir_{k-1}(1-q_k\wir_k)\end{array}\right.
\end{equation}
and the matrix
\begin{equation}\label{dAL1 Vk}
V_k^{(1)}(\bq,\wbr,-h)=\left(
\begin{array}{cc}1-h\lambda^2+hq_k\wir_{k-1} & -h\lambda q_k\\ \\
-h\lambda \wir_{k-1} & 1\end{array}\right).
\end{equation}
and from the Theorem 2 the map 
\begin{equation}\label{dAL-1}
{\cal T}_{-1}(-h):\left\{\begin{array}{l}
(\wiq_k-q_k)/h=-\wiq_{k-1}(1-\wiq_kr_k),\\ \\
(\wir_k-r_k)/h=r_{k+1}(1-\wiq_kr_k)\end{array}\right.
\end{equation}
along with the corresponding matrix 
\begin{equation}\label{dAL-1 Vk}
V_k^{(-1)}(\wbq,\br,-h)=\left(
\begin{array}{cc}1 & h\lambda^{-1}\wiq_{k-1}\\ \\
h\lambda^{-1} r_k & 1+h\lambda^{-2}-h\wiq_{k-1}r_k\end{array}\right).
\end{equation}

{\bf Corollary 1.} {\it Consider the map ${\cal T}_2^{-1}(-h;c):\,
(\bq,\br)\mapsto(\wbq,\wbr)$. This map is defined by the equations of
motion
\begin{equation}
{\cal T}_2^{-1}(-h;c):\left\{\begin{array}{l}
(\wiq_k-q_k)/h=\displaystyle\frac
{(\wiq_{k+2}-\wiq_{k+1}Q_{k+1})\,(1-\wiq_kr_k)}
{(1+h\wiq_{k+1}r_{k-1})}\\ \\
(\wir_k-r_k)/h=-\displaystyle\frac
{(r_{k-2}-r_{k-1}Q_k)\,(1-\wiq_kr_k)}
{(1+h\wiq_{k+1}r_{k-1})}\end{array}\right.
\end{equation}
where
\[
\Big(Q_k+h\wiq_{k+1}r_{k-2}\Big)\,\frac{(1+h\wiq_k\wiq_{k-1}r_{k-1}r_{k-2})\,
(1+h\wiq_{k+1}\wiq_kr_kr_{k-1})}{(1+h\wiq_kr_{k-2})\,(1+h\wiq_{k+1}r_{k-1})},
\]
\[
-\Big(\wiq_{k+1}r_k+\wiq_kr_{k-1}+\wiq_{k-1}r_{k-2}
+h\wiq_{k+1}\wiq_k\wiq_{k-1}r_kr_{k-1}r_{k-2}\Big)=c.
\] 
It approximates the flow {\rm(\ref{AL+2 flow})}
and has a commutation representation {\rm(\ref{d zero curv})} with the role 
of $V_k$ played by the matrix
\begin{equation}\label{dAL2 Wk}
W_k^{(2)}=\frac{1}{1+h\bA_k^{(0)}}
\left(\begin{array}{cc}
1-h\lambda^{2}\wiq_kr_{k-1} & 
-h\lambda^{3}\wiq_{k}-h\lambda \bB_k^{(1)}\\ \\
-h\lambda^{3}r_{k-1}-h\lambda \bC_k^{(1)} & 
1+h\bA_k^{(0)}+h\lambda^{2}\bA_k^{(2)}-h\lambda^4
\end{array}\right).
\end{equation}
Here the quantities $\bA_k^{(2)}$, $\bB_k^{(1)}$, $\bC_k^{(1)}$, $\bA_k^{(0)}$ 
are obtained from the quantities $A_k^{(2)}$, $B_k^{(1)}$, $C_k^{(1)}$, 
$A_k^{(0)}$ by the change $h$ to $-h$, $\bq$ to $\wbq$, and $\wbr$ to $\br$}.

{\bf Corollary 2.} {\it Consider the map ${\cal T}_{-2}^{-1}(-h;c):\,
(\bq,\br)\mapsto(\wbq,\wbr)$. This map is defined by the equations of
motion
\begin{equation}
{\cal T}_{-2}^{-1}(-h;c):\left\{\begin{array}{l}
(\wiq_k-q_k)/h=\displaystyle\frac
{(q_{k-2}-q_{k-1}R_k)\,(1-q_k\wir_k)}
{(1-hq_{k-1}\wir_{k+1})},\\ \\
(\wir_k-r_k)/h=-\displaystyle\frac
{(\wir_{k+2}-\wir_{k+1}R_{k+1})\,(1-q_k\wir_k)}
{(1-hq_{k-1}\wir_{k+1})}\end{array}\right.
\end{equation}
where
\[
\Big(R_k-hq_{k-2}\wir_{k+1}\Big)\,\frac{(1-hq_{k-2}q_{k-1}\wir_{k-1}\wir_k)\,
(1-hq_{k-1}q_k\wir_k\wir_{k+1})}{(1-hq_{k-2}\wir_k)\,(1-hq_{k-1}\wir_{k+1})}
\]
\[
-\Big(q_k\wir_{k+1}+q_{k-1}\wir_k+q_{k-2}\wir_{k-1}-hq_{k-2}q_{k-1}q_k
\wir_{k-1}\wir_k\wir_{k+1}\Big)=c.
\]
It approximates the flow {\rm(\ref{AL-2 flow})}
and has a commutation representation {\rm(\ref{d zero curv})} with the 
role of $V_k$ played by the matrix
\begin{equation}\label{dAL-2 Wk}
W_k^{(-2)}=\frac{1}{1-h\bD_k^{(0)}}
\left(\begin{array}{cc}
1-h\bD_k^{(0)}-h\lambda^{-2}\bD_k^{(-2)}+h\lambda^{-4} & 
h\lambda^{-3}q_{k-1}+h\lambda^{-1}\bB_k^{(-1)}\\ \\
h\lambda^{-3}\wir_k+h\lambda^{-1}\bC_k^{(-1)} &
1+h\lambda^{-2}q_{k-1}\wir_k
\end{array}\right).
\end{equation}
Here the quantities $\bD_k^{(-2)}$, $\bB_k^{(-1)}$, $\bC_k^{(-1)}$, $\bD_k^{(0)}$ 
are obtained from the quantities $D_k^{(-2)}$, $B_k^{(-1)}$, $C_k^{(-1)}$, 
$D_k^{(0)}$ by the change $h$ to $-h$, $\wbq$ to $\bq$, and $\br$ to $\wbr$}.

{\bf Proof.} Since the both Corollaries 1,2 are proved analogously, we 
demonstrate again only the second one. The map ${\cal T}^{-1}_{-2}(-h;c)$ 
obviously allows the commutation representation (\ref{d zero curv}) with the 
matrix $\Big(V_k^{(-2)}(\wbq,\br,-h)\Big)^{-1}$ playing the role of $V_k$.
The Corollary 2 will be proved, if we show that
\[
W_k^{(-2)}=(1+h\lambda^{-4})\,\Big(V_k^{(-2)}(\wbq,\br,-h)\Big)^{-1}.
\]
But from the expressions given in the Theorem 2 it is possible to derive that
\[
\det V_k^{(-2)}(\bq,\wbr,h)=(1-h\lambda^{-4})\,\left(1+hD_k^{(0)}\right),
\]
so that
\[
\det V_k^{(-2)}(\wbq,\br,-h)=(1+h\lambda^{-4})\,\left(1-h\bD_k^{(0)}\right),
\]
which implies the above statement. \qed

\setcounter{equation}{0}
\section{Relation to the Taha--Ablowitz schemes}

From the results of the previous Section it is easy to see that our maps
are indeed some particular cases of the Taha--Ablowitz scheme. Namely:
\begin{itemize}
\item[--] The map ${\cal T}_2(h;c)$  coincides with the Taha--Ablowitz 
difference scheme with the only nonzero parameters $\alpha^{(4)}=1$, 
$\alpha^{(2)}=-c$.
\item[--] The map ${\cal T}_{-2}(h;c)$ coincides with the Taha--Ablowitz 
difference scheme with the only nonzero parameters $\delta^{(-4)}=-1$, 
$\delta^{(-2)}=c$.
\item[--]  The map ${\cal T}_2^{-1}(-h;c)$ coincides with the Taha--Ablowitz 
difference scheme with the only nonzero parameters $\delta^{(4)}=-1$, 
$\delta^{(2)}=c$.
\item[--] The map ${\cal T}_{-2}^{-1}(-h;c)$ coincides with the Taha--Ablowitz 
difference scheme with the only nonzero parameters $\alpha^{(-4)}=1$, 
$\alpha^{(-2)}=-c$.
\end{itemize}

These schemes, obviously, do not satisfy the condition (\ref{dAL2 al=del})
(they certainly should not, because the underlying continuous time flows
(\ref{AL+2 flow}), (\ref{AL-2 flow}) do not allow the reduction (\ref{red})). 
However, certain compositions of our maps, approximating the flow
\begin{equation}\label{AL2 flow}
{\cal F}_2(t)\circ{\cal F}_1(-ct)\circ {\cal F}_{-2}(-t)\circ {\cal F}_{-1}(ct),
\end{equation}
do have the desired property (\ref{dAL2 al=del}), making them suitable
for numerical integration of the space discretized mKdV equation 
(\ref{dMKDV}).

Namely, taking into account that the composition of maps corresponds to the
multiplication of the corresponding $V_k$ matrices, we see immediately that
the following two statements hold.

{\bf Theorem 3.} {\it The map
\begin{equation}\label{dAL2 comp1}
{\cal T}_2(h;c)\circ {\cal T}_{-2}(-h;c)
\end{equation}
approximating the flow {\rm(\ref{AL2 flow})}, coincides with the Taha--Ablowitz
difference scheme with the only nonzero parameters
$\alpha^{(4)}=\delta^{(-4)}=1$, $\alpha^{(2)}=\delta^{(-2)}=-c$, and 
allows the reduction} (\ref{dAL2 red}).

{\bf Theorem 4.} {\it The map
\begin{equation}\label{dAL2 comp2}
{\cal T}_2^{-1}(-h;c)\circ {\cal T}_{-2}^{-1}(h;c)
\end{equation}
approximating the flow {\rm(\ref{AL2 flow})}, coincides with the Taha--Ablowitz
difference scheme with the only nonzero parameters 
$\alpha^{(-4)}=\delta^{(4)}=-1$, $\alpha^{(-2)}=\delta^{(2)}=c$, and 
allows the reduction} (\ref{dAL2 red}).

These two maps (taken by $c=2$) are the first candidates to the role of valid 
difference schemes for numerical integration of the mKdV equation. Just as 
the scheme tested in 
\cite{AT3}, they are integrable, i.e. they have infinitely many integrals of
motion (in the rapidly decaying case; in the $N$--periodic case the number 
of integrals reduces to $N$), and all the integrals are in involution with 
respect to the
Posson bracket (\ref{AL PB}). Additionally, they have an advantage of locality.
More precisely, each step of time integration by the corresponding numerical
scheme consists of solving two local systems of nonlinear equations for the
updates of $\bq$, $\br$. We expect that the savings of the amount of 
computations due to locality will let these schemes to exceed the original
Taha--Ablowitz scheme, even despite the fact that the latter  is of the 
second order in $h$, while (\ref{dAL2 comp1}), (\ref{dAL2 comp2}) are only
of the first order in $h$.

In this connection it should be mentioned that the maps (\ref{dAL2 comp1}), 
(\ref{dAL2 comp2}) are suitable building blocks for applying the 
Ruth--Yoshida--Suzuki techniques \cite{RYS}, which in principle allow to
construct the schemes of an arbitrary high order in $h$. The first step
in this direction is given by the following statement.

{\bf Theorem 5.} {\it The composition
\begin{equation}\label{dAL2 comp3}
{\cal T}_2(h/2;c)\circ {\cal T}_{-2}(-h/2;c)\circ 
{\cal T}_2^{-1}(-h/2;c)\circ {\cal T}_{-2}^{-1}(h/2;c)
\end{equation}
approximates the flow {\rm(\ref{AL2 flow})} up to the second order in $h$.
It coincides with the Taha--Ablowitz difference scheme with the parameters
\[
\alpha^{(4)}=\delta^{(-4)}=-\alpha^{(-4)}=-\delta^{(4)}=\frac{1}{2\Delta},
\]
\[
\alpha^{(2)}=\delta^{(-2)}=-\frac{c(1-h/2)}{2\Delta},\quad
\alpha^{(-2)}=\delta^{(2)}=\frac{c(1+h/2)}{2\Delta},
\]
\[
\alpha^{(0)}=\delta^{(0)}=0,
\]
where
\[
\Delta=1-h^2(1+c^2)/4
\]
In particular, this composition allows the reduction} (\ref{dAL2 red}).

{\bf Proof.} The map (\ref{dAL2 comp3}), as a composition of two 
Taha--Ablowitz schemes, allows a commutation representation (\ref{d zero
curv}) with the matrix $V_k$ whose $A_k$, $D_k$ entries containing, in 
principle, terms with $\lambda^j$, $j=0,\pm 2, \pm 4, \pm 6, \pm 8$, while
the $B_k$, $C_k$ entries contain the terms with $j=\pm 1,\pm 3, \pm 5, \pm 7$.
A careful inspection will convince that only the terms with $|j|\leq 4$ are
actually present in  this matrix. To this end we need the following two
statements.

{\bf Lemma 1.} {\it The entries of the matrix $V_k$ for the map 
{\rm(\ref{dAL2 comp1})} have the following $\lambda$--dependence:}
\begin{eqnarray*}
A(\lambda) & = & 1+h\lambda^4+h\lambda^2\rA_k^{(2)}+h\rA_k^{(0)}+
h\lambda^{-2}\wiq_{k-1}r_k\\
D(\lambda) & = & 1+h\lambda^2q_k\wir_{k-1}+h\rD_k^{(0)}+h\lambda^{-2}
\rD_k^{(-2)}+h\lambda^{-4}\\
B(\lambda) & = & h\lambda^3q_k+h\lambda\rB_k^{(1)}+h\lambda^{-1}\rB_k^{(-1)}+
h\lambda^{-3}\wiq_{k-1}\\
C(\lambda) & = & h\lambda^3\wir_{k-1}+h\lambda\rC_k^{(1)}+h\lambda^{-1}
\rC_k^{(-1)}+h\lambda^{-3}r_k
\end{eqnarray*}

{\bf Lemma 2.} {\it The entries of the matrix $V_k$ for the map 
{\rm(\ref{dAL2 comp1})} have the following $\lambda$--dependence:}
\begin{eqnarray*}
A(\lambda) & = & \Lambda_k\left(1-h\lambda^{2}\wiq_kr_{k-1}+h\rA_k^{(0)}+
h\lambda^{-2}\rA_k^{(-2)}-h\lambda^{-4}\right)\\
D(\lambda) & = & \Lambda_k\left(1-h\lambda^{4}+h\lambda^{2}\rD_k^{(2)}
+h\rD_k^{(0)}-h\lambda^{-2}q_{k-1}\wir_k\right)\\
B(\lambda) & = & \Lambda_k\left(h\lambda^3\wiq_k+h\lambda\rB_k^{(1)}+
h\lambda^{-1}\rB_k^{(-1)}+h\lambda^{-3}q_{k-1}\right)\\
C(\lambda) & = & \Lambda_k\left(h\lambda^3r_{k-1}+h\lambda\rC_k^{(1)}+
h\lambda^{-1}\rC_k^{(-1)}+h\lambda^{-3}\wir_k\right)
\end{eqnarray*}

(The exact expressions for the unspecified functions  
play no role in the following reasonings; therefore we denoted carelessly
different objects in the above statements by one and the same symbol,
such as $\rA_k^{(0)}$). The proof of these two lemmas are similar, so
we give only the {\bf proof of the lemma 1}. Let 
\[
{\cal T}_{-2}(-h;c):\;(\bq,\br)\mapsto(\widehat{\bq},\widehat{\br}),\quad
{\cal T}_2(h;c):\;(\widehat{\bq},\widehat{\br})\mapsto
(\wbq,\wbr)
\]
It follows from the Theorems 1,2 that the the matrix $V_k$ for the composition 
map (\ref{dAL2 comp1}) is equal to
\[
V_k=V_k^{(2)}(\widehat{\bq},\wbr,h)\,V_k^{(-2)}(\widehat{\bq},\br,-h).
\]
The general $\lambda$--dependence of this matrix follows directly from 
(\ref{dAL2 Vk}), (\ref{dAL-2 Vk}), and we have only to prove the explicit
expressions for $\rB_k^{(\pm 3)}$, $\rC_k^{(\pm 3)}$, $\rA_k^{(-2)}$, 
$\rD_k^{(2)}$ given in the formulation of the Lemma. The expressions
$\rC_k^{(3)}=\wir_{k-1}$, $\rC_k^{(-3)}=r_k$ follow immediately, the rest
is derived as follows:
\begin{eqnarray*}
\rB_k^{(3)} & = & \whq_k\left(1-hD_k^{(0)}\right)+hB_k^{(1)}=q_k\\
\rD_k^{(2)} & = & \whq_k\wir_{k-1}\left(1-hD_k^{(0)}\right)+
h\wir_{k-1}B_k^{(-1)}=q_k\wir_{k-1}\\
\rB_k^{(-3)} & = & \whq_{k-1}\left(1-hA_k^{(0)}\right)+hB_k^{(1)}=\wiq_{k-1}\\
\rA_k^{(-2)} & = & \whq_{k-1}r_k\left(1-hA_k^{(0)}\right)+hr_kB_k^{(1)}
=\wiq_{k-1}r_k
\end{eqnarray*}
Here the first two expressions follow from (\ref{dAL-2 proof q}) (of course,
with the change $h$  to $-h$ and $\wbq$ to $\widehat{\bq}$), and the last two
expressions follow from an analogous ''evolution equation for $q_{k-1}$'' for
the map ${\cal T}_2$. \qed

Expressions given in the Lemmas 1,2 imply that in the product
of the corresponding matrices $V_k$ the following terms vanish identically:
the terms with $\lambda^{\pm 8}$, $\lambda^{\pm 6}$ in the entries $A_k$, $D_k$,
the terms with $\lambda^7$, $\lambda^{-5}$ in the entry $B_k$, and the terms
with $\lambda^{-7}$, $\lambda^5$ in the entry $C_k$. After this the commutation
representation (\ref{d zero curv}) shows that the last obstacles, namely the
term with $\lambda^5$ in $B_k$ and the term with $\lambda^{-5}$ in $C_k$,
also must vanish. So, the matrix $V_k$ corresponding to the composition
(\ref{dAL2 comp3})  has the desired $\lambda$--dependence. Now it remains to
observe that the limit values of its entries by $k\to\pm\infty$ are:
\[
\lim_{k\to\pm\infty}A_k(\lambda)=
\Big(1-h\lambda^{-4}/2+h\lambda^{-2}c/2\Big)\,
\Big(1+h\lambda^4/2-h\lambda^2c/2\Big)
\]
\[
=\Delta\,\Big(1+h\lambda^4\alpha^{(4)}+
h\lambda^2\alpha^{(2)}+h\lambda^{-2}\alpha^{(-2)}
+h\lambda^{-4}\alpha^{(-4)}\Big),
\]
\[
\lim_{k\to\pm\infty}D_k(\lambda)=
\Big(1-h\lambda^{4}/2+h\lambda^{2}c/2\Big)\,
\Big(1+h\lambda^{-4}/2-h\lambda^{-2}c/2\Big)
\]
\[
=\Delta\,\Big(1+h\lambda^4\delta^{(4)}+
h\lambda^2\delta^{(2)}+h\lambda^{-2}\delta^{(-2)}
+h\lambda^{-4}\delta^{(-4)}\Big)
\]
with the values $\alpha^{(j)}$, $\delta^{(j)}$ given in the Theorem. \qed

\section{Conclusion}

In the present paper we re-considered the discretizations of the modified
Korteweg-de Vries equation due to Taha and Ablowitz. 
We demonstrated that by some choice of parameters their highly non-local 
scheme may be factorized into the product of much more simple (local) ones, 
each of them approximating a more simple and fundamental flow of the 
Ablowitz--Ladik hierarchy. These local schemes
may be studied exhaustively. In particular, by the same change of variables
as in the previous paper \cite{S5}, we can establish a relation to the pair 
of ''second'' discrete time flows of the relativistic Toda hierarchy, which
gives a simple way to determine interpolating Hamiltonian flows for our
maps and to solve them in terms of a factorization problem in a loop group 
(the relation between the Ablowitz--Ladik and the relativistic Toda hierarchies
is due to \cite{KMZ}).  We guess that also in the
practical computations our variant of the difference scheme will exceed
considerably the old one. It would be interesting and important to carry 
out the corresponding numerical experiments.

The research of the author is financially supported by the DFG (Deutsche
Forschungsgemeinshaft).

\setcounter{equation}{0}
\section{Appendix: \newline difference scheme by Taha and Ablowitz}

The twenty equations, following from (\ref{dAL2 A})--(\ref{dAL2 C}), read:
\begin{eqnarray}
A_{k+1}^{(4)}-A_k^{(4)} & = & 0 
\label{App A4}\\
A_{k+1}^{(-4)}-A_k^{(-4)} & = & \wiq_kC_k^{(-3)}-r_kB_{k+1}^{(-3)}
\label{App A-4}\\ 
A_{k+1}^{(2)}-A_k^{(2)} & = & \wiq_kC_k^{(3)}-r_kB_{k+1}^{(3)}
\label{App A2}\\
A_{k+1}^{(-2)}-A_k^{(-2)} & = & \wiq_kC_k^{(-1)}-r_kB_{k+1}^{(-1)}
\label{App A-2}\\
A_{k+1}^{(0)}-A_k^{(0)} & = & \wiq_kC_k^{(1)}-r_kB_{k+1}^{(1)}
\label{App A0}
\end{eqnarray}
\begin{eqnarray}
D_{k+1}^{(-4)}-D_k^{(-4)} & = & 0
\label{App D-4}\\
D_{k+1}^{(4)}-D_k^{(4)} & = & \wir_kB_k^{(3)}-q_kC_{k+1}^{(3)}
\label{App D4}\\
D_{k+1}^{(-2)}-D_k^{(-2)} & = & \wir_kB_k^{(-3)}-q_kC_{k+1}^{(-3)}
\label{App D-2}\\
D_{k+1}^{(2)}-D_k^{(2)} & = & \wir_kB_k^{(1)}-q_kC_{k+1}^{(1)}
\label{App D2}\\
D_{k+1}^{(0)}-D_k^{(0)} & = & \wir_kB_k^{(-1)}-q_kC_{k+1}^{(-1)}
\label{App D0}
\end{eqnarray}
\begin{eqnarray}
B_k^{(3)} & = & q_kA_{k+1}^{(4)}-\wiq_kD_k^{(4)}
\label{App B3}\\
B_{k+1}^{(-3)} & = & \wiq_kD_k^{(-4)}-q_kA_{k+1}^{(-4)}
\label{App B-3}\\
B_k^{(1)} & = & q_kA_{k+1}^{(2)}-\wiq_kD_k^{(2)}+B_{k+1}^{(3)}
\label{App B1}\\
B_{k+1}^{(-1)} & = & \wiq_kD_k^{(-2)}-q_kA_{k+1}^{(-2)}+B_k^{(-3)}
\label{App B-1}
\end{eqnarray}
\begin{eqnarray}
C_k^{(-3)} & = & r_kD_{k+1}^{(-4)}-\wir_kA_k^{(-4)}
\label{App C-3}\\
C_{k+1}^{(3)} & = & \wir_kA_k^{(4)}-r_kD_{k+1}^{(4)}
\label{App C3}\\
C_k^{(-1)} & = & r_kD_{k+1}^{(-2)}-\wir_kA_k^{(-2)}+C_{k+1}^{(-3)}
\label{App C-1}\\
C_{k+1}^{(1)} & = & \wir_kA_k^{(2)}-r_kD_{k+1}^{(2)}+C_k^{(3)}
\label{App C1}
\end{eqnarray}
\begin{eqnarray}
\wiq_k-q_k & = & 
h\Big(q_kA_{k+1}^{(0)}-\wiq_kD_k^{(0)}+B_{k+1}^{(1)}-B_k^{(-1)}\Big)
\label{App ev q}\\
\wir_k-r_k & = & 
h\Big(r_kD_{k+1}^{(0)}-\wir_kA_k^{(0)}+C_{k+1}^{(-1)}-C_k^{(1)}\Big)
\label{App ev r}
\end{eqnarray}

Taha and Ablowitz interpret the first 18 equations 
(\ref{App A4})--(\ref{App C1}) as the defining relations for the 18 
coefficients $A_k^{(j)}$--$D_k^{(j)}$, and the last two equations 
(\ref{App ev q}), (\ref{App ev r}) as the evolution equations for
$q_k$, $r_k$. This leads to a highly nonlocal scheme. Let us indicate 
how do Taha and Ablowitz solve the first 18 equations (with some amendments).

First of all, (\ref{App A4}), (\ref{App D-4}) imply
\begin{equation}\label{App A4D-4 res}
A_k^{(4)}=\alpha^{(4)},\qquad D_k^{(-4)}=\delta^{(-4)}
\end{equation}
where $\alpha^{(4)}$, $\delta^{(-4)}$ are some constants (and the same holds 
for other $\alpha^{(j)}$, $\delta^{(j)}$ appearing further on). Substituting
(\ref{App B3}), (\ref{App C3}) into (\ref{App D4}), and (\ref{App B-3}),
(\ref{App C-3}) into (\ref{App A-4}), we get upon use of (\ref{App A4D-4 res})
the following difference equations:
\begin{equation}
D_{k+1}^{(4)}(1-q_kr_k)=D_k^{(4)}(1-\wiq_k\wir_k), \quad
A_{k+1}^{(-4)}(1-q_kr_k)=A_k^{(-4)}(1-\wiq_k\wir_k).
\end{equation}
Their solution may be put into the form
\begin{equation}\label{App D4A-4 res}
D_k^{(4)}=\delta^{(4)}\Lambda_k,\qquad A_k^{(-4)}=\alpha^{(-4)}\Lambda_k,
\end{equation}
where
\begin{equation}\label{App Lamk}
\Lambda_k=\prod_{j=-\infty}^{k-1} \frac{1-\wiq_j\wir_j}{1-q_jr_j}
\end{equation}
is for all $k$ of order $1+O(h)$ and tends to 1 as $k\to\pm\infty$ (due to the
infinite-dimensional analog of the integral of motion (\ref{AL H0})).

The back substitution (\ref{App A4D-4 res}), (\ref{App D4A-4 res}) into
(\ref{App B3}), (\ref{App C3}), (\ref{App B-3}), (\ref{App C-3}) results in
\begin{equation}\label{App B3 res}
B_k^{(3)}=\alpha^{(4)}q_k-\delta^{(4)}\wiq_k\Lambda_k,
\end{equation}
\begin{equation}\label{App C-3 res}
C_k^{(-3)}=\delta^{(-4)}r_k-\alpha^{(-4)}\wir_k\Lambda_k,
\end{equation}
\begin{equation}\label{App B-3 res}
B_k^{(-3)}=\delta^{(-4)}\wiq_{k-1}-\alpha^{(-4)}q_{k-1}\Lambda_k,
\end{equation}
\begin{equation}\label{App C3 res}
C_k^{(3)}=\alpha^{(4)}\wir_{k-1}-\delta^{(4)}r_{k-1}\Lambda_k.
\end{equation}
Substituting the last formulas into (\ref{App A2}), (\ref{App D-2}), 
we arrive at the following difference equations:
\begin{equation}\label{App A2 dif}
A_{k+1}^{(2)}-A_k^{(2)}=\alpha^{(4)}(\wiq_k\wir_{k-1}-q_{k+1}r_k)+
\delta^{(4)}(\wiq_{k+1}r_k\Lambda_{k+1}-\wiq_kr_{k-1}\Lambda_k),
\end{equation}
\begin{equation}\label{App D-2 dif}
D_{k+1}^{(-2)}-D_k^{(-2)}=\delta^{(-4)}(\wiq_{k-1}\wir_k-q_kr_{k+1})+
\alpha^{(-4)}(q_k\wir_{k+1}\Lambda_{k+1}-q_{k-1}\wir_k\Lambda_k).
\end{equation}
Their solutions are given by
\begin{equation}\label{App A2 res}
A_k^{(2)}=\alpha^{(2)}-\alpha^{(4)}(q_kr_{k-1}+\Phi_k)+
\delta^{(4)}\wiq_kr_{k-1}\Lambda_k,
\end{equation}
\begin{equation}\label{App D-2 res}
D_k^{(-2)}=\delta^{(-2)}-\delta^{(-4)}(q_{k-1}r_k+\Psi_k)+
\alpha^{(-4)}q_{k-1}\wir_k\Lambda_k,
\end{equation}
where the quantites
\begin{equation}\label{App PhiPsik}
\Phi_k = \sum_{j=-\infty}^{k-1} (q_jr_{j-1}-\wiq_j\wir_{j-1}),\quad
\Psi_k = \sum_{j=-\infty}^{k-1} (q_{j-1}r_j-\wiq_{j-1}\wir_j)
\end{equation}
are of order $O(h)$ and vanish by $k\to\pm\infty$ (the statement for 
$k\to+\infty$ follows from the fact that the infinite dimensional analogs
of the quantities $H_{\pm 1}(\bq,\br)$ from (\ref{AL H pm1}) are the integrals
of motion).

To proceed further, we substitute (\ref{App B1}), (\ref{App C1}) into
(\ref{App D2}). Using (\ref{App B3 res}), (\ref{App C3 res}), and 
(\ref{App A2 dif}), we derive from (\ref{App D2}) the following 
difference equation:
\[
(D_{k+1}^{(2)}-\alpha^{(4)}q_{k+1}\wir_k)(1-q_kr_k)-
(D_k^{(2)}-\alpha^{(4)}q_k\wir_{k-1})(1-\wiq_k\wir_k)=
\]
\[
-\delta^{(4)}\Big(\wiq_{k+1}\wir_k(1-q_kr_k)\Lambda_{k+1}-
q_kr_{k-1}(1-\wiq_k\wir_k)\Lambda_k\Big).
\]
Analogously, substituting (\ref{App B-1}), (\ref{App C-1}) into (\ref{App A-2})
and using (\ref{App B-3 res}), (\ref{App C-3 res}), (\ref{App D-2 dif}), we 
obtain:
\[
(A_{k+1}^{(-2)}-\delta^{(-4)}\wiq_kr_{k+1})(1-q_kr_k)-
(A_k^{(-2)}-\delta^{(-4)}\wiq_{k-1}r_k)(1-\wiq_k\wir_k)=
\]
\[
-\alpha^{(-4)}\Big(\wiq_k\wir_{k+1}(1-q_kr_k)\Lambda_{k+1}-
q_{k-1}r_k(1-\wiq_k\wir_k)\Lambda_k\Big).
\]
A little trick (apparently missed in \cite{AT3}), based on the formula
$(1-q_kr_k)\Lambda_{k+1}=(1-\wiq_k\wir_k)\Lambda_k$, allows to solve
these difference equations as 
\begin{equation}\label{App D2 res}
D_k^{(2)}=\delta^{(2)}\Lambda_k+\alpha^{(4)}q_k\wir_{k-1}-\delta^{(4)}
(\wiq_k\wir_{k-1}-\Phi_k)\Lambda_k,
\end{equation}
\begin{equation}\label{App A-2 res}
A_k^{(-2)}=\alpha^{(-2)}\Lambda_k+\delta^{(-4)}\wiq_{k-1}r_k-\alpha^{(-4)}
(\wiq_{k-1}\wir_k-\Psi_k)\Lambda_k.
\end{equation}

Now we have determined all the quantities from the right--hand 
sides of (\ref{App B1}), (\ref{App B-1}), (\ref{App C1}), (\ref{App C-1}).
Collecting all the results, we find:
\begin{eqnarray}
B_k^{(1)} & = & \alpha^{(2)}q_k-\delta^{(2)}\wiq_k\Lambda_k
\nonumber\\
&&+\alpha^{(4)}\Big(q_{k+1}-q_k(q_{k+1}r_k+q_kr_{k-1}+\Phi_k)\Big)
\nonumber\\
&&-\delta^{(4)}\Big(\wiq_{k+1}-\wiq_k(\wiq_{k+1}\wir_k+\wiq_k\wir_{k-1}-
\Phi_k)\Big)\Lambda_k
\label{App B1 res}\\
B_k^{(-1)} & = & \delta^{(-2)}\wiq_{k-1}-\alpha^{(-2)}q_{k-1}\Lambda_k
\nonumber\\
&&+\delta^{(-4)}\Big(\wiq_{k-2}-\wiq_{k-1}(q_{k-1}r_k+\wiq_{k-2}\wir_{k-1}+
\Psi_k)\Big)\nonumber\\
&& -\alpha^{(-4)}\Big(q_{k-2}-q_{k-1}(q_{k-2}r_{k-1}+\wiq_{k-1}\wir_k-\Psi_k)
\Big)\Lambda_k\label{App B-1 res}\\
C_k^{(1)} & = & \alpha^{(2)}\wir_{k-1}-\delta^{(2)}r_{k-1}\Lambda_k
\nonumber\\
&&+\alpha^{(4)}\Big(\wir_{k-2}-\wir_{k-1}(q_kr_{k-1}+\wiq_{k-1}\wir_{k-2}+\Phi_k)
\Big)\nonumber\\
&&-\delta^{(4)}\Big(r_{k-2}-r_{k-1}(q_{k-1}r_{k-2}+\wiq_k\wir_{k-1}-\Phi_k)\Big)
\Lambda_k
\label{App C1 res}\\
C_k^{(-1)} & = & \delta^{(-2)}r_k-\alpha^{(-2)}\wir_k\Lambda_k
\nonumber\\
&&+\delta^{(-4)}\Big(r_{k+1}-r_k(q_kr_{k+1}+q_{k-1}r_k+\Psi_k)\Big)
\nonumber\\
&&-\alpha^{(-4)}\Big(\wir_{k+1}-\wir_k(\wiq_k\wir_{k+1}+\wiq_{k-1}\wir_k-\Psi_k)
\Big)\Lambda_k
\label{App C-1 res}
\end{eqnarray}

Finally, these expressions, being substituted in (\ref{App A0}), (\ref{App D0}),
result in difference equations whose solutions may be represented as
\[
A_k^{(0)} = \alpha^{(0)}-\alpha^{(2)}(q_kr_{k-1}+\Phi_k)
+\delta^{(2)}\wiq_kr_{k-1}\Lambda_k
\]
\[
-\alpha^{(4)}(q_{k+1}r_{k-1}+q_kr_{k-2}-\wiq_{k+1}q_k\wir_kr_{k-1}-
\wiq_kq_k\wir_{k-1}r_{k-1}-\wiq_{k}q_{k-1}\wir_{k-1}r_{k-2}+\Xi_k)
\]
\begin{equation}\label{App A0 res}
+\delta^{(4)}\Big(\wiq_{k+1}r_{k-1}+\wiq_kr_{k-2}-\wiq_kr_{k-1}
(\wiq_{k+1}\wir_k+\wiq_kr_{k-1}+q_{k-1}r_{k-2}-\Phi_k)\Big)\Lambda_k
\end{equation}
\[
\]
\[
D_k^{(0)} = \delta^{(0)}-\delta^{(-2)}(q_{k-1}r_k+\Psi_k)
+\alpha^{(-2)}q_{k-1}\wir_k\Lambda_k
\]
\[
-\delta^{(-4)}(q_{k-1}r_{k+1}+q_{k-2}r_k-\wiq_kq_{k-1}\wir_{k+1}r_k-
\wiq_{k-1}q_{k-1}\wir_kr_k-\wiq_{k-1}q_{k-2}\wir_{k}r_{k-1}+\Omega_k)
\]
\begin{equation}\label{App D0 res}
+\alpha^{(-4)}\Big(q_{k-1}\wir_{k+1}+q_{k-2}\wir_k-q_{k-1}\wir_k
(\wiq_k\wir_{k+1}+q_{k-1}\wir_k+q_{k-2}r_{k-1}-\Psi_k)\Big)\Lambda_k
\end{equation}
Here
\[
\Xi_k=\sum_{j=-\infty}^{k-1}\Big(q_jr_{j-2}-\wiq_j\wir_{j-2}
-q_{j+1}r_j\Phi_{j+3}+\wiq_j\wir_{j-1}\Phi_{j-1}\Big)
\]
\[
\Omega_k=\sum_{j=-\infty}^{k-1}\Big(q_{j-2}r_j-\wiq_{j-2}\wir_j
-q_jr_{j+1}\Psi_{j+3}+\wiq_{j-1}\wir_{j}\Psi_{j-1}\Big)
\]
These quantities, just as $\Phi_k$, $\Psi_k$, are of order $O(h)$ and vanish 
in both limits $k\to\pm\infty$. The see that this is true for $k\to +\infty$,
one has to use the partial summation. For example, for $\Xi_k$ one arrives
at the following expression:
\[
\Xi_k=q_kr_{k-1}\Phi_{k+2}-\frac{1}{2}(q_kr_{k-1}\phi_{k-1}+
\wiq_k\wir_{k-1}\widetilde{\phi}_{k-1}-2\wiq_k\wir_{k-1}\phi_{k-1})
\]
\[
+\sum_{j=-\infty}^{k-1}\Big(q_jr_{j-2}-\frac{1}{2}q_jr_{j-1}
(q_{j+1}r_j+q_jr_{j-1}+q_{j-1}r_{j-2})\Big)
\]
\[
-\sum_{j=-\infty}^{k-1}\Big(\wiq_j\wir_{j-2}-\frac{1}{2}\wiq_j\wir_{j-1}
(\wiq_{j+1}\wir_j+\wiq_j\wir_{j-1}+\wiq_{j-1}\wir_{j-2})\Big),
\]
where
\[
\phi_k=\sum_{j=-\infty}^{k-1}q_jr_{j-1},\quad{\rm so}\;\;{\rm that}\quad
\Phi_k=\phi_k-\widetilde{\phi}_k.
\]
The obtained expression for $\Xi_k$ clearly vanishes at $k\to +\infty$ due
to the fact that an infinite-dimensional analog of (\ref{AL H2}) is an
integral of motion.

Now the Taha--Ablowitz difference scheme for the mKdV is given by
(\ref{App ev q}), (\ref{App ev r}) with the expressions (\ref{App B1 res})--
(\ref{App D0 res}). Due to the appearance of the quantities $\Lambda_k$,
$\Phi_k$, $\Psi_k$, $\Xi_k$, $\Omega_k$ this scheme is not only implicit,
but also highly non-local.  More precisely, they consider only the reduced case
$\br=\pm\bq$, assuming $\delta^{(j)}=\alpha^{(-j)}$. In this reduction
$A_k^{(j)}=D_k^{(-j)}$, $B_k^{(j)}=\pm C_k^{(-j)}$, and, of course,
$\Phi_k=\Psi_k$, $\Xi_k=\Omega_k$, so that the number of the nonlocal
ingredients is reduced by two.

\end{document}